# Optimized design of a Penning ion source for sealed neutron tube

Jia Song[a], Shengduo Liu[a], Weiyang Zhang[a], Sijia Zhou[a], Hailong Xu[a], Zebin Li[a], Tian Zhang[a], Zhihua Gao[a], Shiwei Jing [a,b,*], Guofeng Qu[a,*]

[a] *School of Physics, Northeast Normal University, Changchun, 130024, China*

[b] *Neutron Bless Radiation (Jilin)High-Tech company, Changchun, 130024, , China*

*Corresponding author: E-mail address: jingsw504@nenu.edu.cn

**Abstract:** Sealed neutron tubes have a wide range of applications, and the ion source is their core component. Penning ion sources commonly suffer from issues such as uneven magnetic field distribution and a low proportion of monoatomic ions. Improving the performance of the ion source can effectively address the problems of low neutron flux and short operational lifespan. This study aims to optimise the magnetic field configuration and discharge parameters of the ion source, thereby increasing the proportion of monoatomic and enhancing discharge stability, and to provide a design basis for high-performance sealed neutron tubes. Develop a magnetic field-plasma coupling model to compare and analyze the magnetic field distribution patterns of traditional magnetic block structures and soft iron-reinforced structures, and investigate the mechanisms by which operating pressure and anode voltage affect plasma density and ion composition using COMSOL multiphysics simulation methods. Simulation results indicate that the soft iron structure significantly enhances the axial magnetic field strength and uniformity within the discharge region; under conditions of 0.06 Pa gas pressure and 1500 V anode voltage, the proportion of monoatomic ions increased from the conventional 9% to 30%.

## 1. Introduction

As a small gas pedal neutron source, compact neutron generator has the advantages of small size, compact structure, convenient storage and transportation, and controllable yield [1]. Compared with the ordinary isotope neutron source, the gas pedal neutron source has the advantages of high neutron yield, good monochromaticity of the energy spectrum, and easy shutdown, which has successfully expanded the neutron applications from large-scale laboratories to the on-site scenarios. For example, the use of neutron-sensitized $\gamma$-rays for elemental logging in oil and gas exploration [2], the realization of rapid safety inspection of air containers through transmission imaging [3], and the real-time analysis of material composition in industrial online monitoring [4].In addition to traditional industrial and security applications, miniaturized neutron sources have promising applications in neutron irradiation testing of fusion reactor materials, neutron photography, etc., which are of great significance to the study of small neutron tubes.

When applying neutron tubes to the demanding scenarios described above, their performance still faces a number of bottlenecks. First, its neutron yield is relatively limited compared with that of large sources, resulting in a low detection signal-to-noise ratio and a long measurement time required. Second, its neutron energy spectrum is relatively homogeneous, which is not adaptable enough for applications requiring thermal neutrons or neutrons in specific energy regions. In addition, the lifetime and stability of neutron tubes under long-term operation are key constraints to their deployment in high-intensity, continuous industrial environments. Together, these shortcomings have resulted in neutron tube-based systems having upper limits in detection sensitivity, accuracy and reliability, making it difficult to meet the growing demand for high-end applications.

Penning ion sources with the advantages of simple structure and easy operation are widely used in ion gas pedals [5], ion implantation and fusion applications. The performance bottleneck of neutron tubes can be largely attributed to the performance limitations of their core component, the ion source. The history of the Penning ion source can be traced back to the beginning of the 20th century. In 1936, Frans Miehel Penning invented the Penning discharge. In the working mechanism of the neutron tube, the ion source is responsible for generating and inducing a beam of deuterium ions for bombarding a tritium target. The beam intensity of the ion source directly determines the deuterium-tritium reaction rate, i.e., the neutron yield; the beam quality, the elicitation efficiency, and the plasma density and stability affect the stability of the neutron output as well as the lifetime of the ion source itself [6]. Common optimization methods for neutron tube ion sources include: selecting cathode and absorber materials with low sputtering rate and high bombardment resistance to radically extend the working life of the ion source [7]; improving the plasma density and uniformity by improving the magnetic field configuration or the discharge mode [8-10] to improve the beam intensity; designing better electrode geometry [11] and potential distribution to reduce the emission of the ion beam; distribution to reduce ion beam divergence and loss and improve the elicitation efficiency [12].

Currently, ion sources operating under vacuum and low-pressure discharges face two problems: first, the thermogenic demagnetization of permanent magnets under unfavorable conditions of vacuum heat dissipation [12], which leads to magnetic field degradation and unstable performance; and second, the discharge parameters affecting the plasma chemical processes, which results in the proportion of monatomic ions generally lower than 10 percent [13], thus decreasing the yield of neutrons. Based on this, this study proposes a set of optimization schemes aimed at synergistically optimizing the above problems. First, in terms of magnetic field stability, we introduce a soft iron based on the conventional permanent magnet structure. The design aims to delay the working demagnetization phenomenon of the permanent magnet and enhance the stability of the permanent magnet. The magnetic field line distribution is optimized to form a more uniform and stronger axial magnetic field in the discharge region to enhance the plasma discharge efficiency. Second, in terms of improving the monatomic ion ratio, we focus on the two adjustable parameters that are most direct to the plasma generation process: gas pressure and anode cylinder voltage. Through the synergistic optimization of the gas pressure and voltage parameters, the plasma state can be regulated to the most favorable region for the generation of hydrogen ions. To this end, this study will adopt the finite element analysis method by building magnetic field and plasma simulation models in order to reveal the intrinsic physical mechanism, which will be useful for the development of a high-performance long-life neutron tube ion source.

## 2. Principles and Methods

### 2.1 Working principle

A neutron tube is a typical vacuum device, with its tube body providing the required vacuum environment and meeting high-voltage insulation requirements. Common tube body materials include ceramics and glass. The neutron tube primarily consists of four components: the target, the acceleration system, the ion source, and the gas pressure regulation system, as shown in **Fig. 1**. Its operating principle involves the ion source ionizing the neutral gas inside the tube under external circuit control to generate plasma. The accelerated ion beam, either focused or diverged by the acceleration system, strikes the target. There, it reacts with deuterium and tritium isotopes to produce neutrons. Ultimately, the generated neutron beam is effectively utilized across diverse fields, ranging from fundamental scientific research to practical engineering applications.

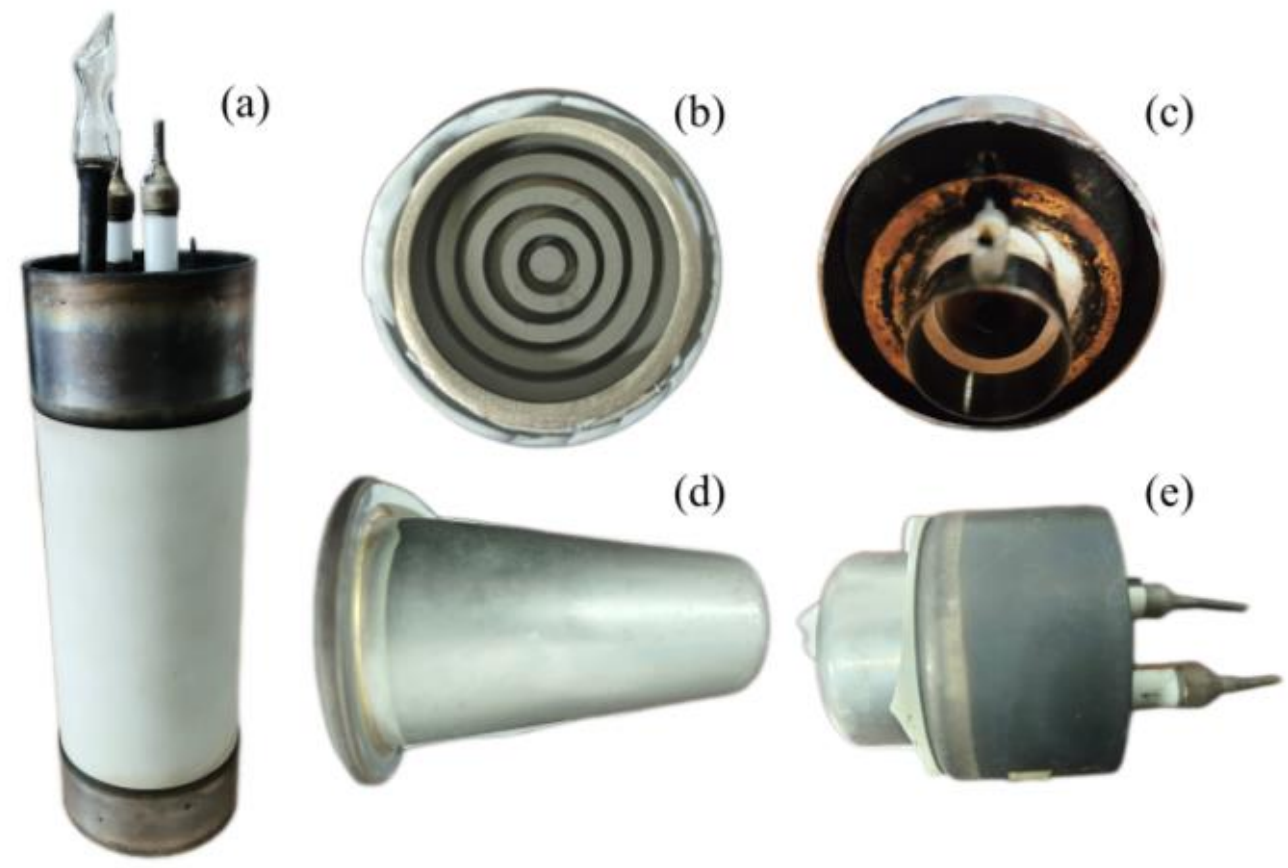

**Fig. 1** (a) Neutron tube physical diagram, (b) Target, (c) Ion source, (d) Acceleration system, (e) Anode shield.

A standard Penning ion source features two cathodes, a hollow cylindrical anode, and a permanent magnet; it operates via a DC voltage applied between the cathodes and anode [14]. The symmetrical electrode arrangement increases the ionization electron path length, thereby enhancing discharge efficiency. Upon ionization initiation, electrons oscillate axially several times within the cylinder between the cathodes. Collisions between electrons and neutral gas molecules are the primary mechanism for producing charged particles, including electrons and positive ions. In addition to axial drift, the applied axial magnetic field forces electrons into helical trajectories along the field lines. This spiraling motion confines them to these paths, effectively suppressing radial diffusion toward the anode wall [15]. The trajectories of ionized electrons follow magnetic field lines while undergoing Larmor precession, with a characteristic Larmor radius given by:$R = (m_e v_\perp) / (e \cdot B)$. Where: $m_e$ denotes the electron mass, B represents the axial magnetic field, and $v_\perp$ is the velocity component perpendicular to the axial direction. This turning radius R is commonly termed the Larmor radius. The magnetic field within the ion source effectively confines the compressed plasma density. Ion density increases with magnetic field strength, while the radial motion of electrons slows according to magnetic field intensity. This effectively prolongs electron residence time and increases the number of ionization collisions, thereby enhancing plasma density. Reducing particle losses decreases the input energy required for the system to generate and sustain plasma at a specified density [17]. Penning ion sources allow for ion extraction along two paths: axial extraction through the cathode aperture or radial extraction through a split in the anode oriented perpendicular to the axis.

### 2.2 Plasma theoretical model

In this study, a numerical simulation method based on multi-physics field coupling is adopted, aiming to implement and analyze the internal magnetic field distribution of the ion source, the behavioral process of gas discharge plasma [18][19]. Electron behavior was modeled via drift-diffusion equations for density and mean energy [20], with electron convection neglected. Regarding fundamental details of electron transport,

$$\frac{\partial}{\partial t}(n_e)+\nabla\cdot[-n_e(\mu_e\cdot \mathrm{E})-D_e\cdot\nabla n_e]=S_e \tag{1}$$

$$\frac{\partial}{\partial t}(n_\varepsilon)+\nabla\cdot[-n_\varepsilon(\mu_\varepsilon\cdot \mathrm{E})-D_\varepsilon\cdot\nabla n_\varepsilon]+\mathrm{E}\cdot\Gamma_{\mathrm{e}}=S_\varepsilon \tag{2}$$

where the electron source $S_e=\sum_{j=1}^{M}x_jk_jN_nn_e$ , $x_j$ is the mole fraction of the target substance for reaction j, $K_j$ is the rate coefficient of reaction j, while $N_n$ is the total neutral number density. The net electron energy loss due to inelastic collisions is the sum of the energy losses from all individual reactions, $S_\varepsilon=\sum_{j=1}^{P}x_jk_jN_nn_e\Delta\varepsilon_j$.

Incorporating the electron energy distribution function (EEDF) [21], the rate coefficient r is obtained:

$$k=\gamma\int_0^\infty \varepsilon f(\varepsilon)\sigma(\varepsilon)d\varepsilon \tag{3}$$

where $\gamma=\left(\frac{2q}{m_e}\right)^{\frac{1}{2}}$ , $m_e$ is the electron, ε is the energy, $\sigma k$ is the collision cross section, $f$ is the electron energy distribution function, and $f(\varepsilon)$ is the non-Maxwellian EEDF, which can be calculated by resolving the Boltzmann equations [22]. The electron energy distribution function used in this study is taken from reference [23], while $\sigma(\varepsilon)$ is the corresponding collision cross section. The dominant processes in the Penning plasma can be summarized by the 12 reactions presented in **Table 1** [11].

**Table 1** The main reaction in the process of PIG hydrogen discharge.

| No. | Reactions | Description | Rate coefficients |
|---|---|---|---|
| 1 | $e+H_2\rightarrow 2H+e$ | Dissociation | $\alpha_1$ |
| 2 | $H+H+\mathrm{wall}\rightarrow H_2+wall$ | H wall recombination | $\alpha_2$ |
| 3 | $e+H\rightarrow H^++2e$ | H ionization | $\alpha_3$ |
| 4 | $e+H_2^+\rightarrow H^++H+e$ | Dissociative excitation | $\alpha_4$ |
| 5 | $e+H_2\rightarrow H_2^++2e$ | Molecular ionization | $\alpha_5$ |
| 6 | $e+H_2^+\rightarrow H+H^*$ | Dissociative recombination | $\alpha_6$ |
| 7 | $H_2^+ + H_2 \rightarrow H_3^+ + H$ | $H_3^+$ ion formation | $\alpha_7$ |
| 8 | $e + H_3^+ \rightarrow 2H + H^+ + e$ | Dissociative excitation | $\alpha_8$ |
| 9 | $e + H_3^+ \rightarrow 3H$ | Dissociative recombination | $\alpha_9$ |
| 10 | $e + H_2 \rightarrow H^+ + H + 2e$ | Dissociative ionization | $\alpha_{10}$ |
| 11 | $e + H_2^+ \rightarrow 2H^+ + 2e$ | Dissociative ionization | $\alpha_{11}$ |
| 12 | $e + H_2^+ \rightarrow H^+ + H^* + e$ | Dissociative excitation | $\alpha_{12}$ |

**Fig. 2** presents the calculated volumetric reaction rate coefficients $\sigma(\varepsilon)$. Here, T denotes electron temperature for electronic reactions and ion temperature for heavy-particle reactions, the latter coefficients are evaluated relative to their target particles.

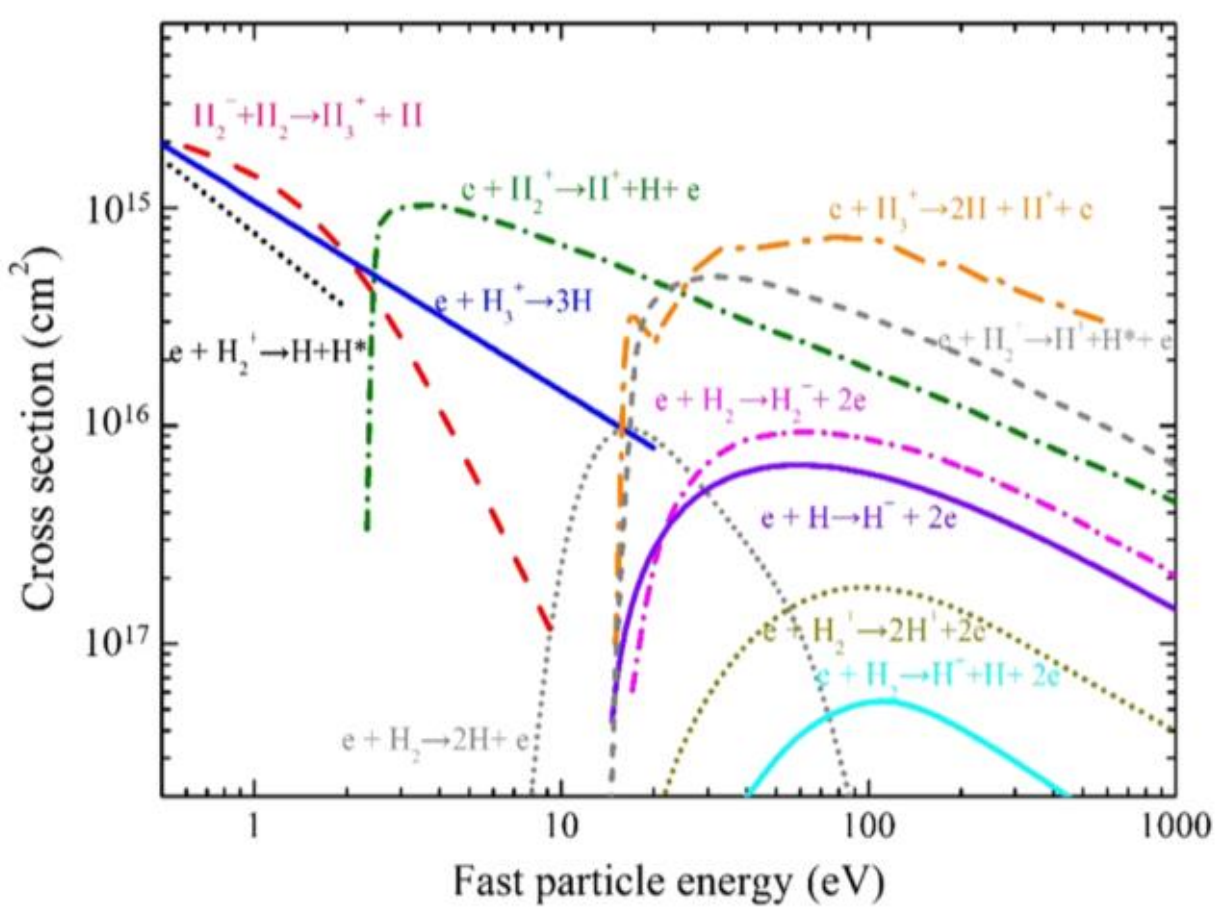

**Fig. 2** Schematic of the dominant processes within the hydrogen plasma of the ion source[24].

## 3. Results and Analysis

### 3.1 Magnetic field design

The magnetic field plays a critical role in Penning ion sources. Previous studies have shown that although multi-magnetic-ring structures can enhance the efficiency of Penning ion sources, they lead to excessive source size[10]. To improve the uniformity of the axial magnetic field, this study optimized the magnet configuration by introducing a soft iron structure. The ion source is modeled within an infinite air domain, with the core configuration of its magnetic block structure detailed in **Fig. 3(a)**. The magnetic field is provided by a samarium cobalt rare-earth magnet placed along the axial direction with a magnet size of Φ 16 × 4 mm, and **Fig. 3(b)** depicts the resulting magnetic field distribution.

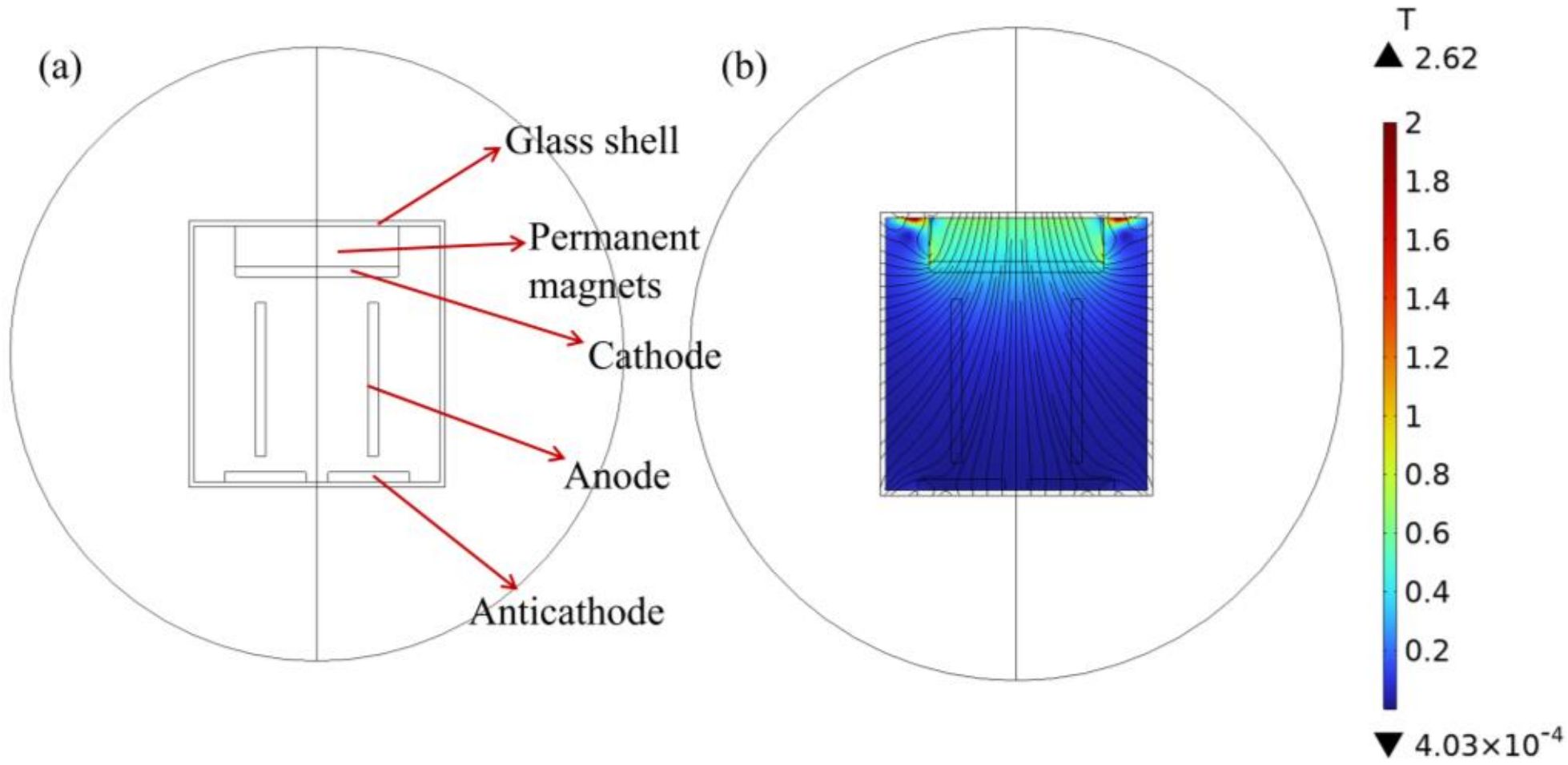

**Fig. 3** Magnetic block-type ion source: (a) two-dimensional model; (b) distribution of magnetic flux density and field lines within the source.

For the purpose of optimizing the magnetic field distribution with greater efficiency, a Penning ion source was designed by adding a soft iron ring around the periphery of the pair of cathodes on the basis of a magnetic block-type ion source, which was used to adjust the magnetic field line distribution, all other conditions being unchanged. Effective optimization of the magnetic field is achieved by adjusting the dimensions and material of the soft iron. The primary structure of this soft-iron-type ion source is shown in **Fig. 4(a)**. Comparison of its magnetic field characteristics with **Fig. 3** shows that the distribution of magnetic induction lines inside the soft-iron type structure **Fig. 4(b)** is significantly denser.

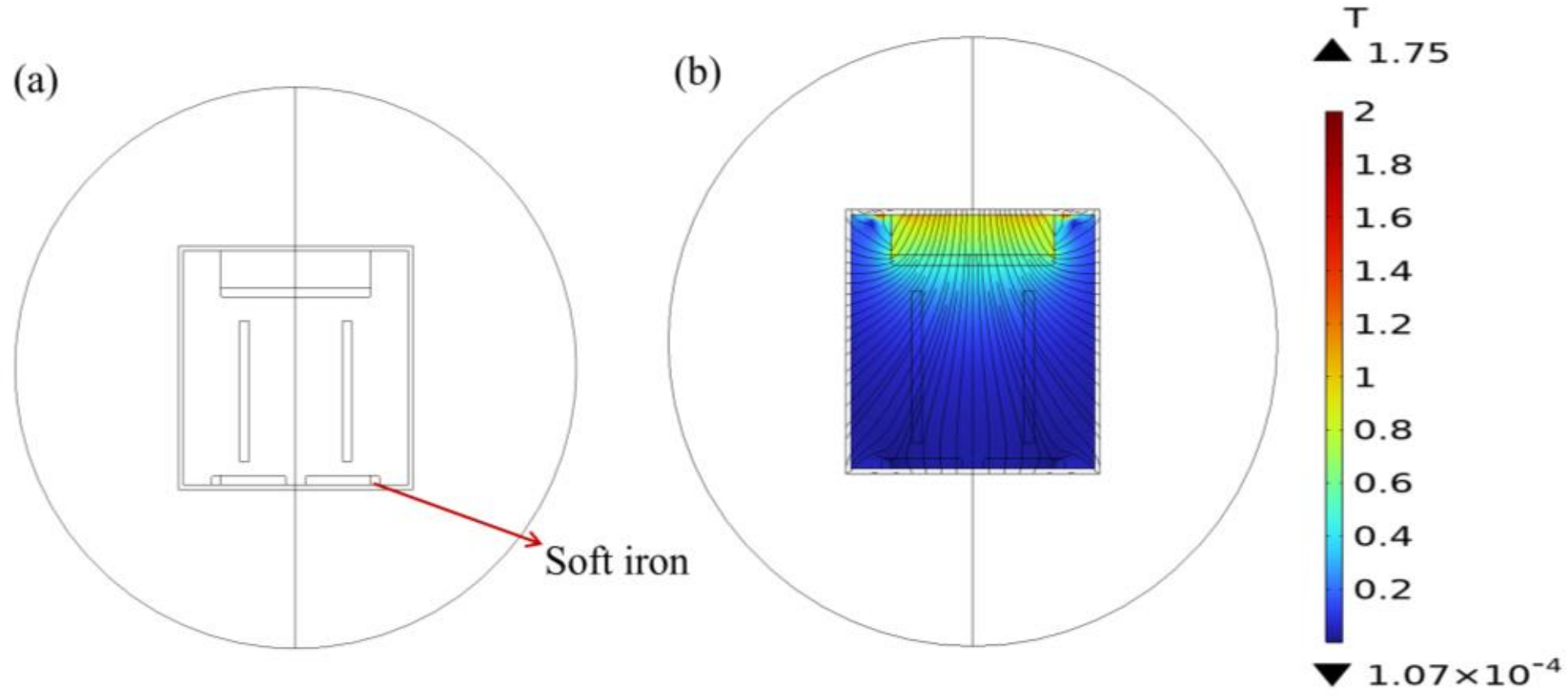

**Fig. 4** Soft-iron-type ion source: (a) Two-dimensional geometric model; (b) Distribution of magnetic flux density and corresponding field lines within the model.

The two-dimensional magnetic field distribution cloud map clearly demonstrates the magnetic field differences between the two structures. To further compare the magnetic field distribution patterns of magnetic blocks and soft iron, this section selects characteristic cross-sections along the axial and radial directions of the discharge chamber, with their intersection point serving as the origin of the chamber as shown in **Figure 5**. By extracting magnetic induction intensity data at these cross-sections and plotting distribution curves, we conduct quantitative analysis of the axial and radial distribution disparities between the two structures.

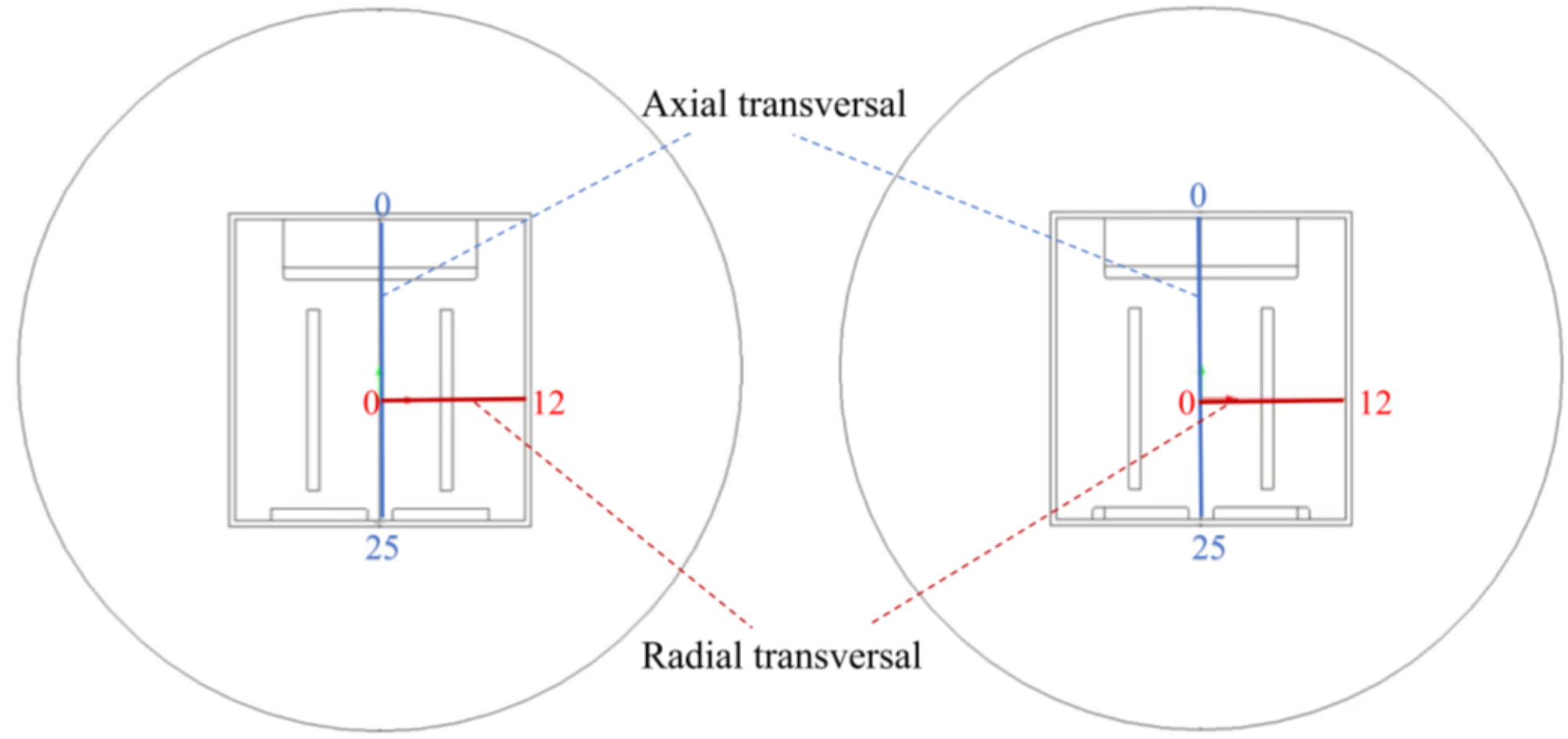

**Fig. 5** Location and length of the intercept line.

The axial magnetic field intensity distributions of the two ion source designs are compared, as shown in **Fig. 6**. Although both exhibit decay along the axial, the soft-iron-enhanced design consistently maintains a higher field strength than the configuration without soft iron. The magnetic field strength with soft iron decreases slower with increasing axial distance indicating that the soft iron has a polymagnetization effect, which enhances and maintains the magnetic field; the magnetic field strength without soft iron decreases faster with increasing distance indicating that the magnetic field decays more significantly in the absence of soft iron.

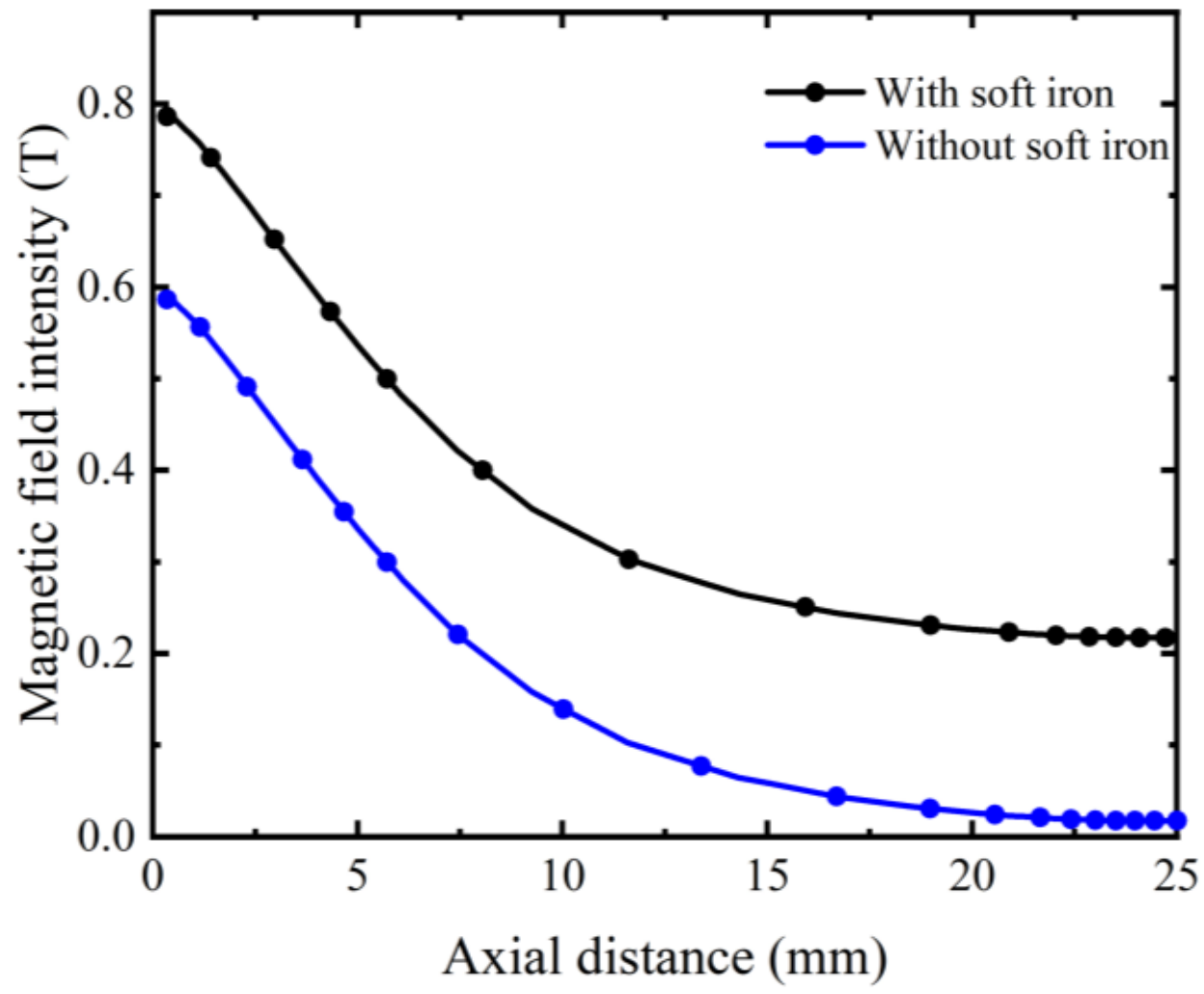

**Fig. 6** Axial magnetic field comparison at 1 T surface flux.

The radial magnetic field intensity distributions of these two ion sources are compared in **Fig. 7**. While both exhibit a reduction near the center, their fields remain relatively uniform radially, with the soft-iron design consistently showing higher strength. Consequently, combined with the axial analysis in **Fig. 6**, the soft-iron-type ion source demonstrates a highly uniform magnetic field distribution that fully satisfies the design requirements.

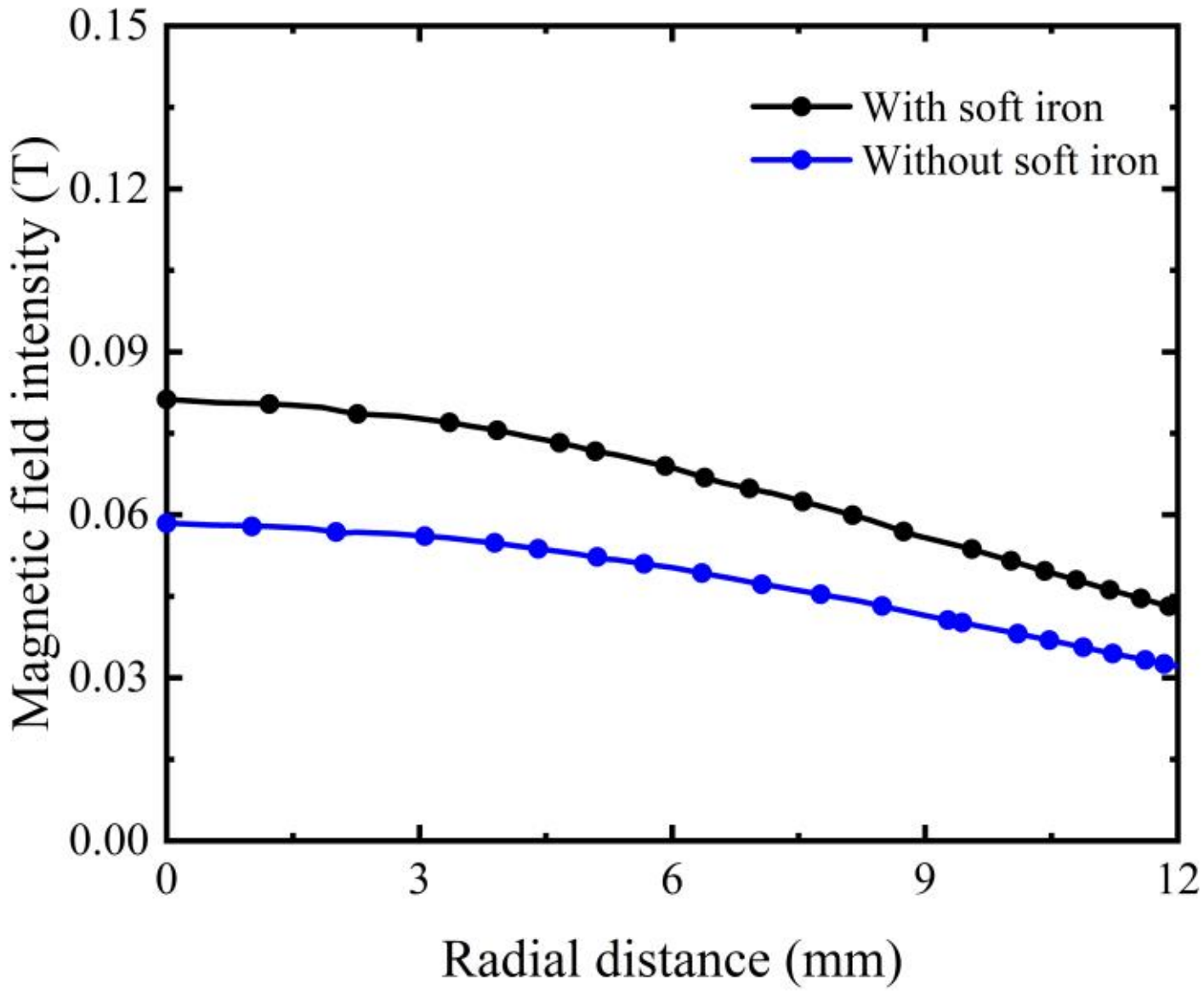

**Fig. 7** Radial magnetic field comparison at 1 T surface flux.

### 3.2 Plasma optimization

To facilitate the coupled simulation of the magnetic field and plasma while improving computational efficiency, this study employs a two-dimensional axisymmetric model to investigate the discharge region in detail. The model gives a schematic structure of the Penning source discharge region as in Fig. 8.The discharge region is mainly composed of permanent magnet, anode cylinder, cathode plate, anticathode plate, soft iron and glass shell.

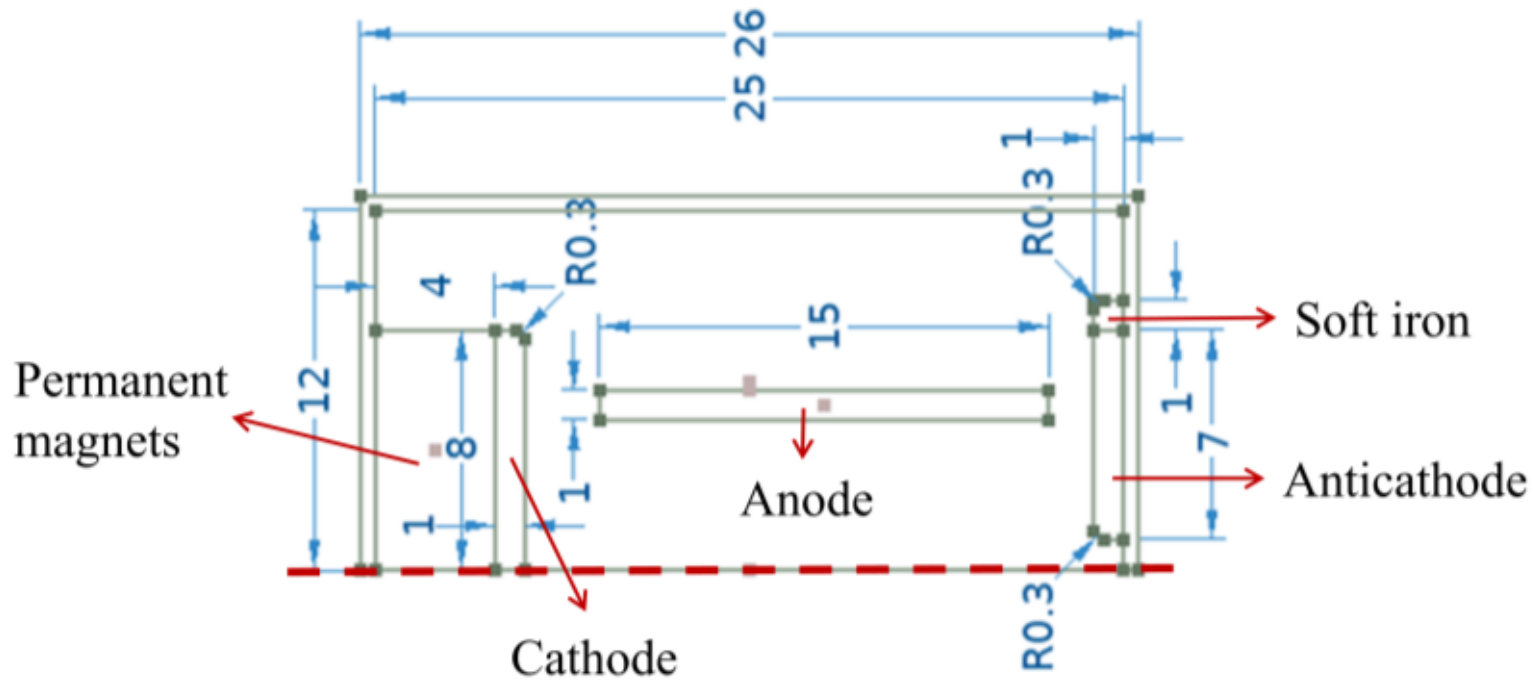

**Fig. 8** Two-dimensional axisymmetric model of the ion source, with the symmetry axis marked by a red dashed line.

The cathode and counter-cathode of the ion source are positioned along the axial direction and are both grounded. A central axial lead hole (approximately 2 mm in diameter) in the counter-cathode connects to the neutron tube's lead system for axial ion extraction. The anode cylinder, with dimensions of 12 mm in diameter and 15 mm in height, is stabilized with an applied voltage of 1500 V. The anode cylinder, cathode and counter cathode materials are copper. The magnet size is Φ16 × 4mm, and the soft iron ring size is Φ(16-18) ×1mm.The discharge area is filled with 0.12 Pa of hydrogen, and the blue area shown in **Fig. 9** is the plasma simulation area.

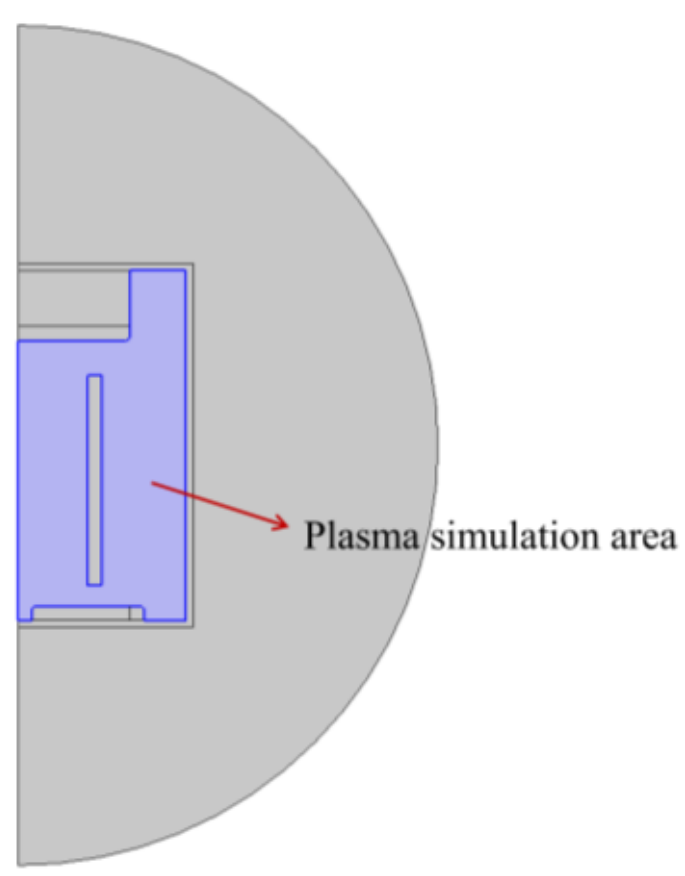

**Fig. 9** Plasma simulation region filled with 0.12 Pa hydrogen gas.

Under the exemplary conditions of 1500 V anode voltage and 0.12 Pa gas pressure at steady state, **Fig. 10** presents the spatial distributions of electron and ion densities, utilizing mirror symmetry to fully represent the 2D axisymmetric structure. To facilitate observation of the discharge region, the color legend was adjusted, including a redefinition of its upper limit. As shown in **Fig. 10(a)**, the highest electron density is concentrated at the plasma axis center, correlating with the region of stronger magnetic field. The ion density from neutral gas ionization is concentrated near the lead hole, with $H^+$ and $H_2^+$ being negligible elsewhere. Their detailed distributions within the ion source are specifically shown in **Fig. 10(b)** and **(c)**.

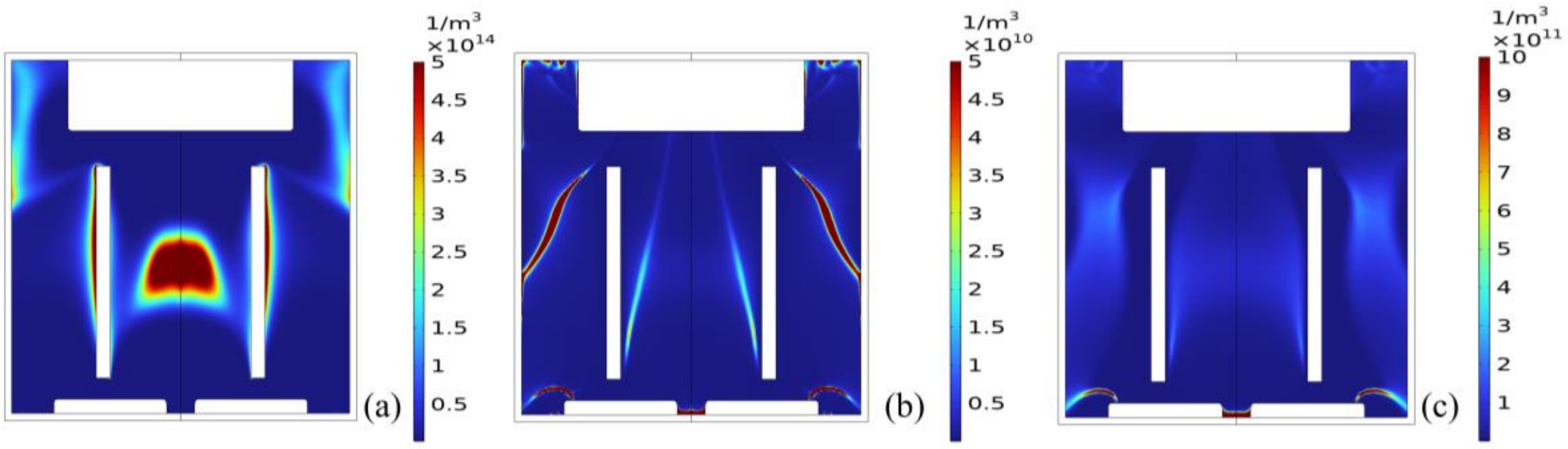

**Fig. 10** Particle density distributions at 1500 V and 0.12 Pa: (a) electrons, (b) $H^+$, (c) $H_2^+$.

This work investigates the axial and radial distributions of electron and ion densities, presented in **Fig. 11**. Axially, the electron density peaks within the central plasma discharge region, indicating an area of high ionization completeness, before decreasing **Fig. 11(a)**. In contrast, the radial electron density profile declines slowly, then rapidly, followed by a slight rebound **Fig. 11(b)**. For hydrogen ions , the axial density initially stabilizes and then peaks at approximately 17 mm, a location near the lead hole characterized by high ion density **Fig. 11(c)**.In **Fig. 11(d)**, with the increase of radial distance, the ion number density first increases to reach the peak and then decreases. From **Fig. 11(e)**, it can be observed that with the increase of axial distance, the ion number density starts to decrease after some fluctuation. **Fig. 11(f)** Number density along radial distance increases slowly to peak and then decreases rapidly.

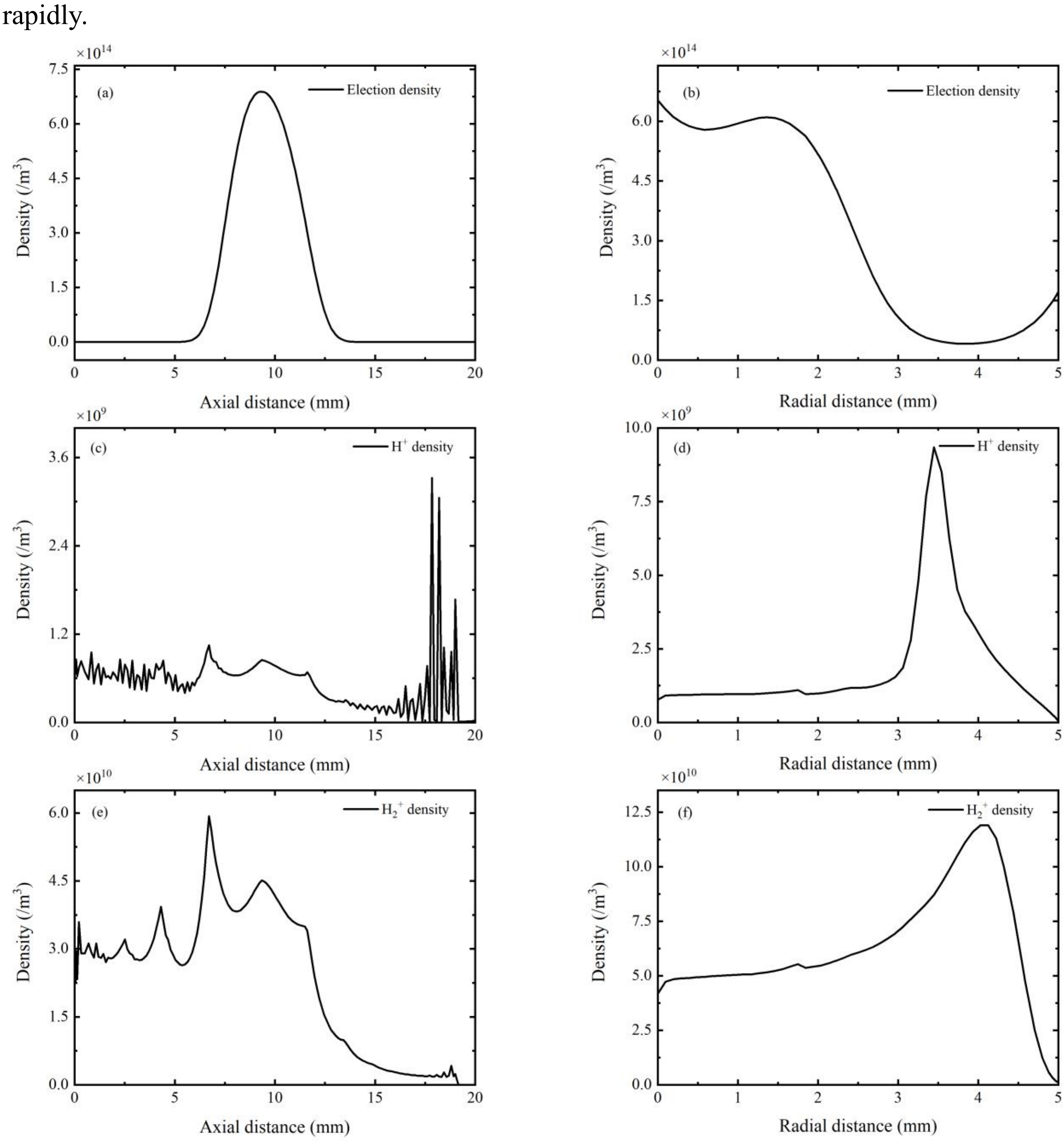

**Fig. 11** Count density distributions for electrons, $H^+$, and $H_2^+$ along the axial and radial profile lines: (a, b) electrons, (c, d) $H^+$, and (e, f) $H_2^+$.

This study aims to determine the distribution of electrons, $H^+$, and $H_2^+$ under varying gas pressures. Under the premise of a self-sustained Penning discharge, simulations were conducted at five gas pressures (0.06, 0.12, 0.3, 0.5, and 1 Pa) to investigate their effects on electron and ion number densities. To isolate the influence of pressure, the anode cylinder voltage was fixed at 1500 V.

Electrons are the lightest and fastest-moving particles in a plasma, and they serve as the fundamental driving force behind the discharge. They absorb energy from the electric field, transfer that energy to other particles through collisions, and drive chemical reactions. They determine the intensity, ionization degree, and stability of the discharge. **Figure 12** shows the profile distribution of electron number density at different gas pressures. The distribution patterns of electron number density in the axial and radial directions are shown in **Figure 13**. When the hydrogen pressure increases from 0.06 Pa to 0.12 Pa, the peak of the axial electron number density rises, and the radial electron density also shows an upward trend. When the pressure continues to rise to 1 Pa, the peak of the electron number density begins to decline, and the radial distance corresponding to the peak in the fluctuation zone shifts to the right as the pressure increases.

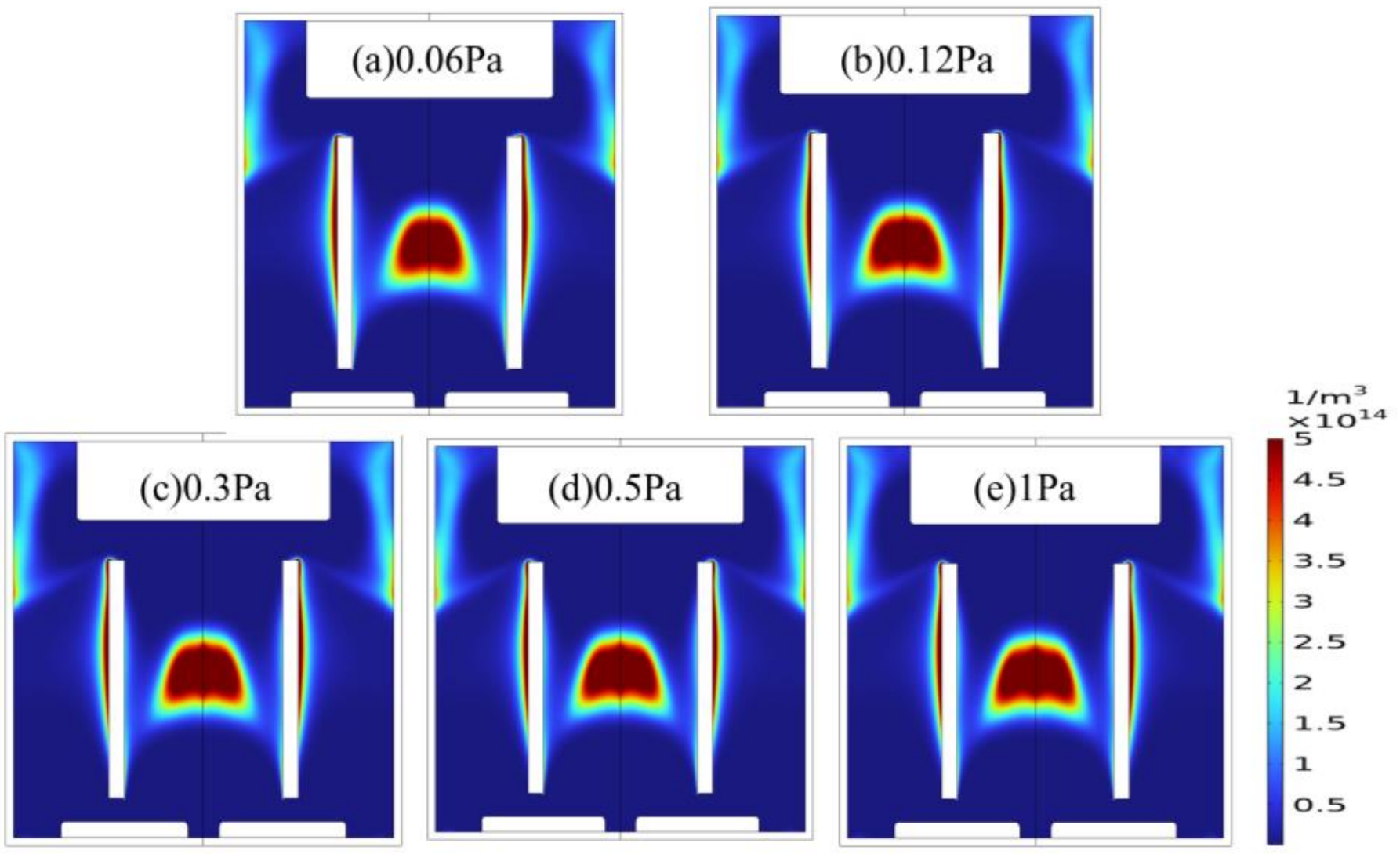

**Fig. 12** Electron number density distribution map.

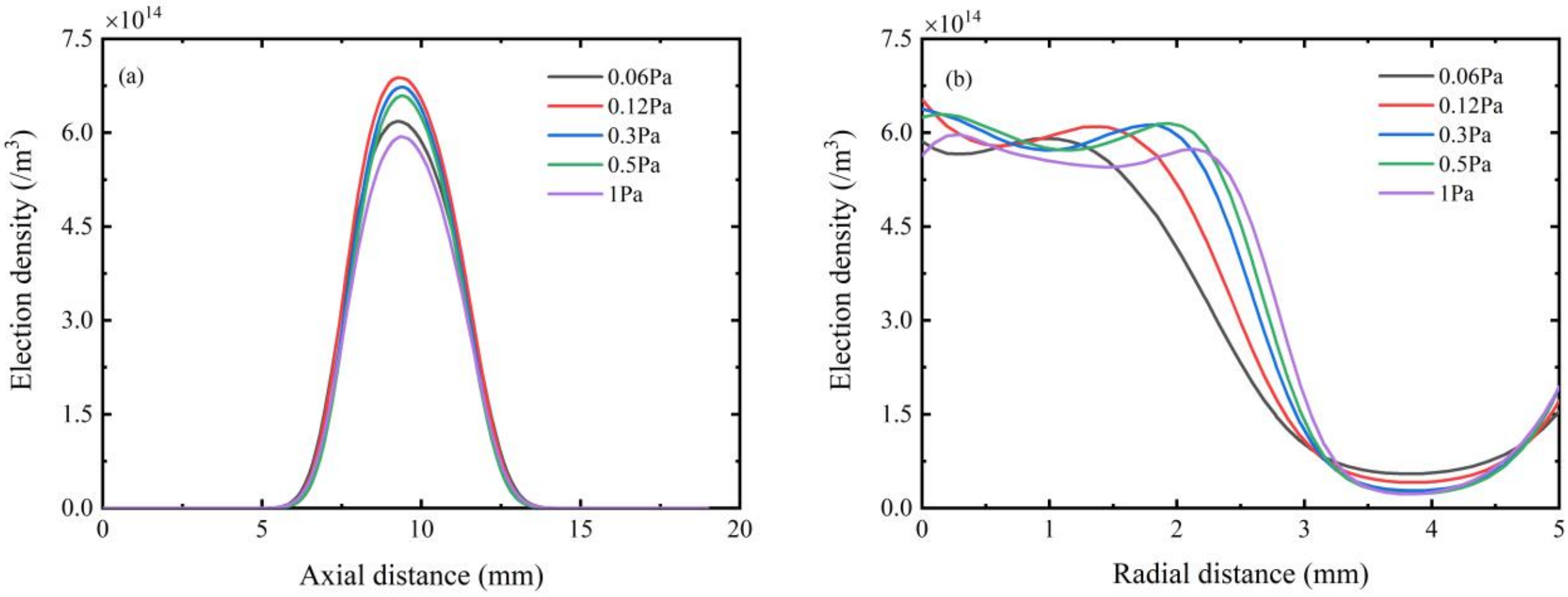

**Fig. 13** Distribution pattern of electron number density in axial and radial directions.

The hydrogen ion serves as an indicator of high-energy discharge and a harbinger of chemical reactivity; it signifies the presence of high-energy electrons and a high-density hydrogen environment, and exhibits extremely high chemical reactivity. It is the dominant ionic component in high-intensity discharges and is key to the generation of high-energy proton beams. Figures 14 and 15 show, respectively, the overall variation of the hydrogen ion number density with gas pressure and its spatial distribution in the axial and radial directions. As the background gas pressure in the reaction chamber increases, the distribution of hydrogen ion density in the plasma undergoes systematic changes. When the hydrogen gas pressure rises from 0.06 Pa to 0.5 Pa, the axial hydrogen atom ion number density increases accordingly, while the radial electron density also shows an upward trend and shifts toward the anode cylinder. When the pressure continues to rise to 1 Pa, the hydrogen ion number density shows a decreasing trend in both the axial and radial directions.

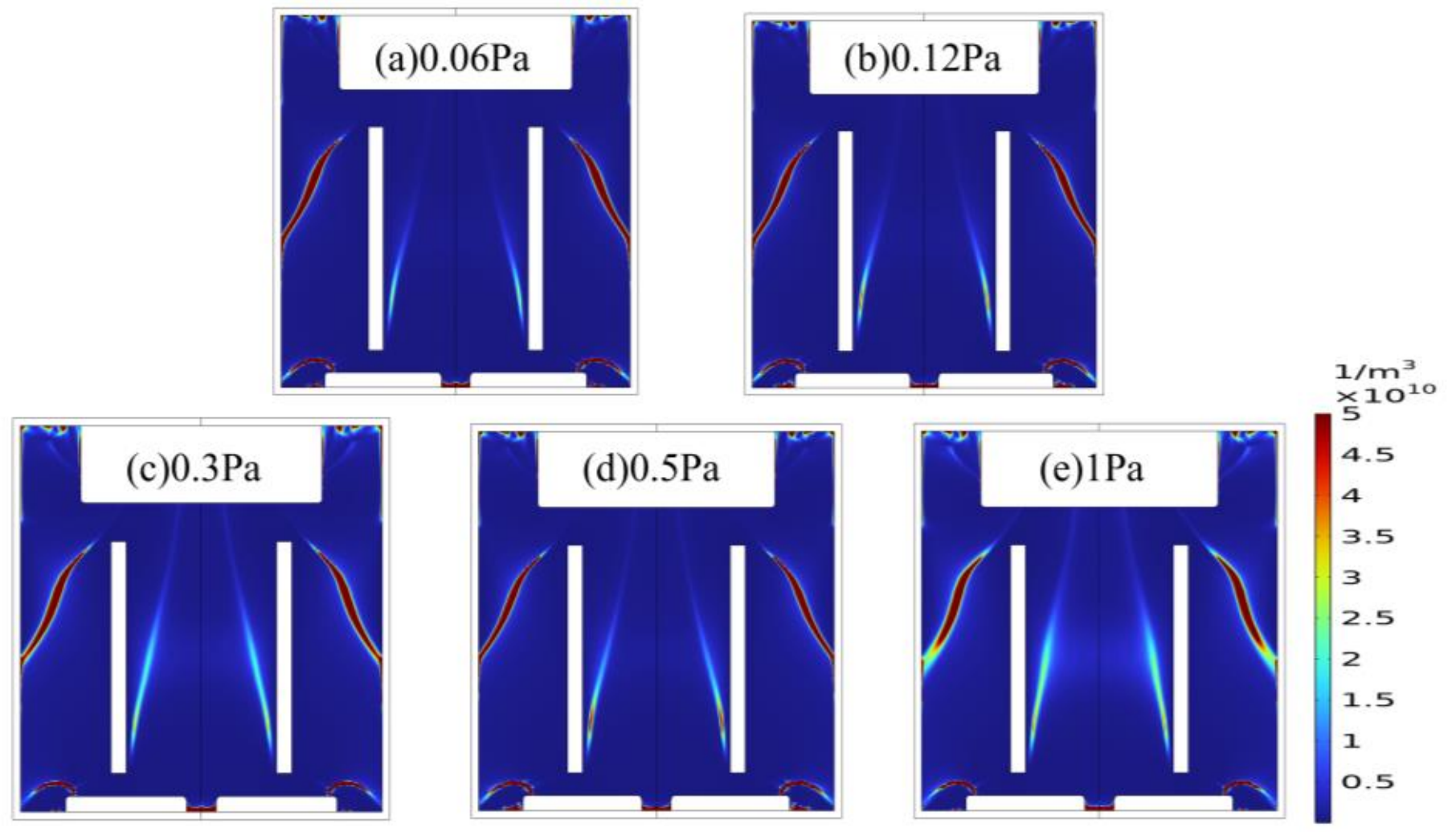

**Fig. 14** $H^+$ number density distribution map.

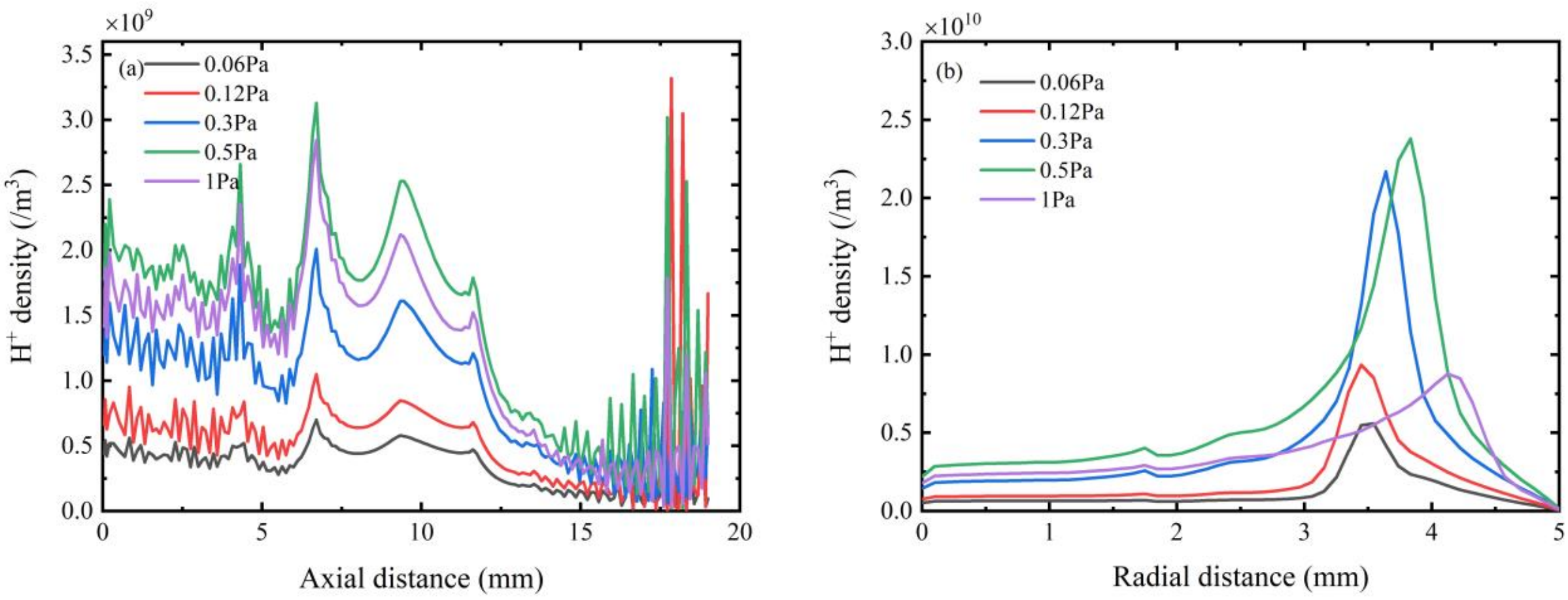

**Fig. 15** Distribution pattern of $H^+$ number density in axial and radial directions.

Hydrogen molecular ions are the predominant ions during the initial stages of discharge, generating a large number of hydrogen atoms through dissociation and recombination. They are the dominant ionic species in weak-to-moderate-intensity discharges and serve as the primary source of hydrogen atomic gas. **Figure 16** shows the distribution of the number density of hydrogen molecular ions in the ion source as a function of gas pressure. **Figure 17** presents the distribution of hydrogen molecular ion number density along the axial and radial directions. As shown in Figures 16 and 17, the axial hydrogen molecular ion number density increases as the gas pressure rises from 0.06 Pa to 0.5 Pa; however, when the pressure continues to rise to 1 Pa, the hydrogen molecular ion number density begins to decrease. At the same time, changes in gas pressure likely alter the equilibrium conditions within the system. This change manifests specifically as follows: in the radial dimension, the peak position of the ion density is influenced by gas pressure; as the pressure increases, the position corresponding to the peak shifts toward the anode plate.

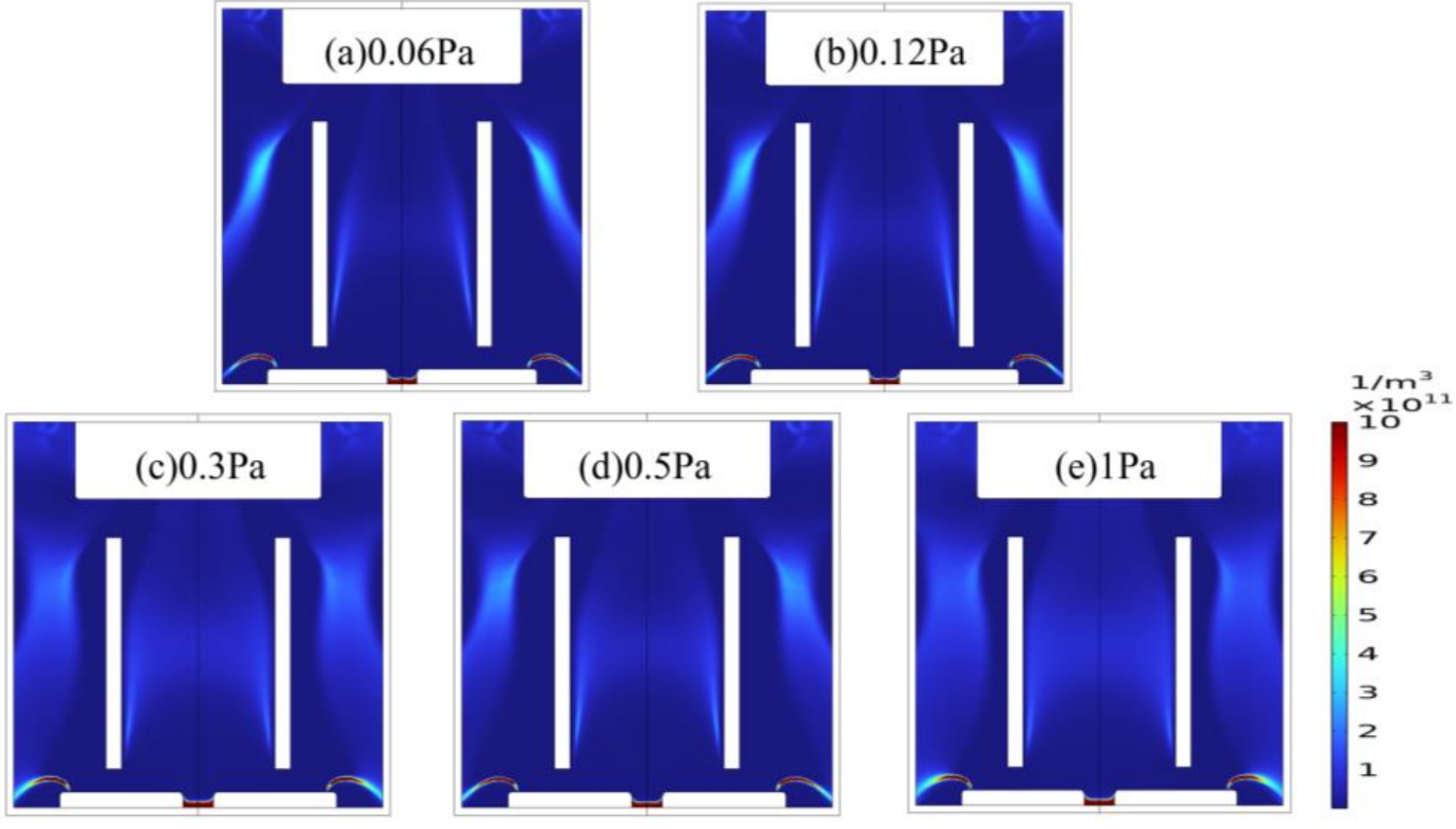

**Fig. 16** $H_2^+$ number density distribution map.

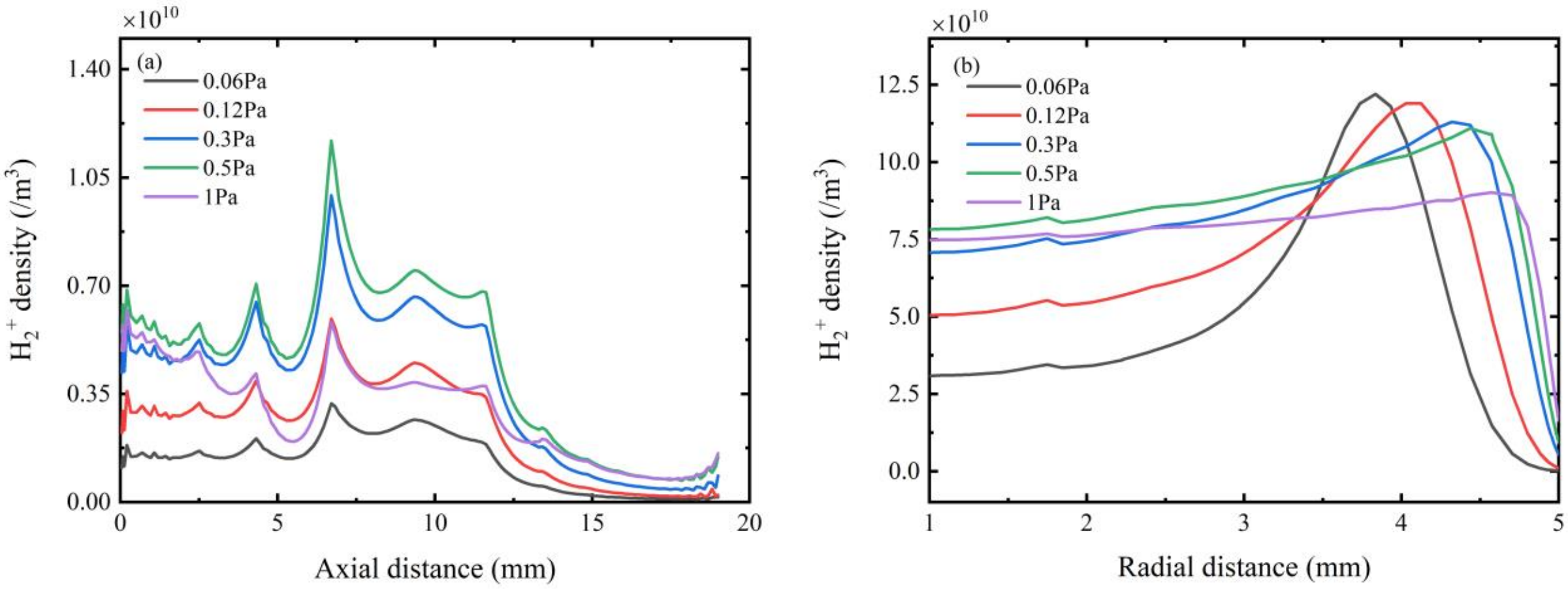

**Fig. 17** Distribution pattern of $H_2^+$ number density in axial and radial directions.

The voltage across the anode cylinder is also a critical parameter affecting the discharge of a neutron tube Penning ion source. Drawing on extensive design experience with Penning-type neutron tubes, simulation experiments investigating the effect of voltage variations across the anode cylinder on the electron and ion number densities primarily examined three voltage levels 800 V, 1000 V, 1200 V, 1400 V, and 1600 V, with the aim of identifying the distribution patterns of electrons, hydrogen ions, and hydrogen atom ions at different voltages. Therefore, the discharge pressure was fixed at 0.5 Pa. The profile distribution of electron number density at different voltages is shown in **Figure 18**. The distribution patterns of electron number density in the axial and radial directions are shown in **Figure 19**. As the voltage increases from 800 V to 1600 V along the axial direction, the peak electron number density gradually increases, and the trend of the curve remains consistent; the higher the voltage, the higher the electron number density at the same radial distance, and the trend of the curve remains consistent.

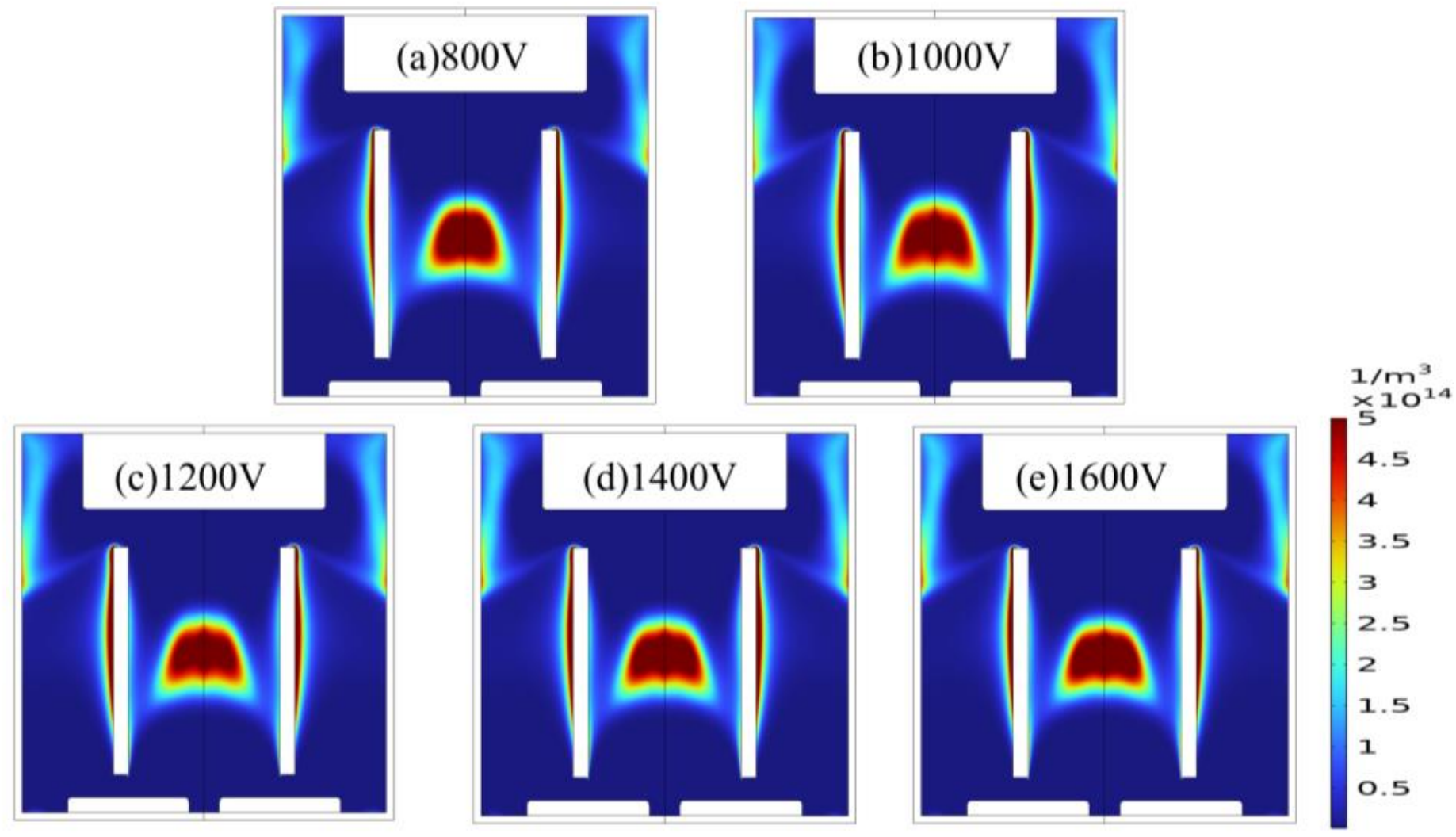

**Fig. 18** Electron number density distribution map.

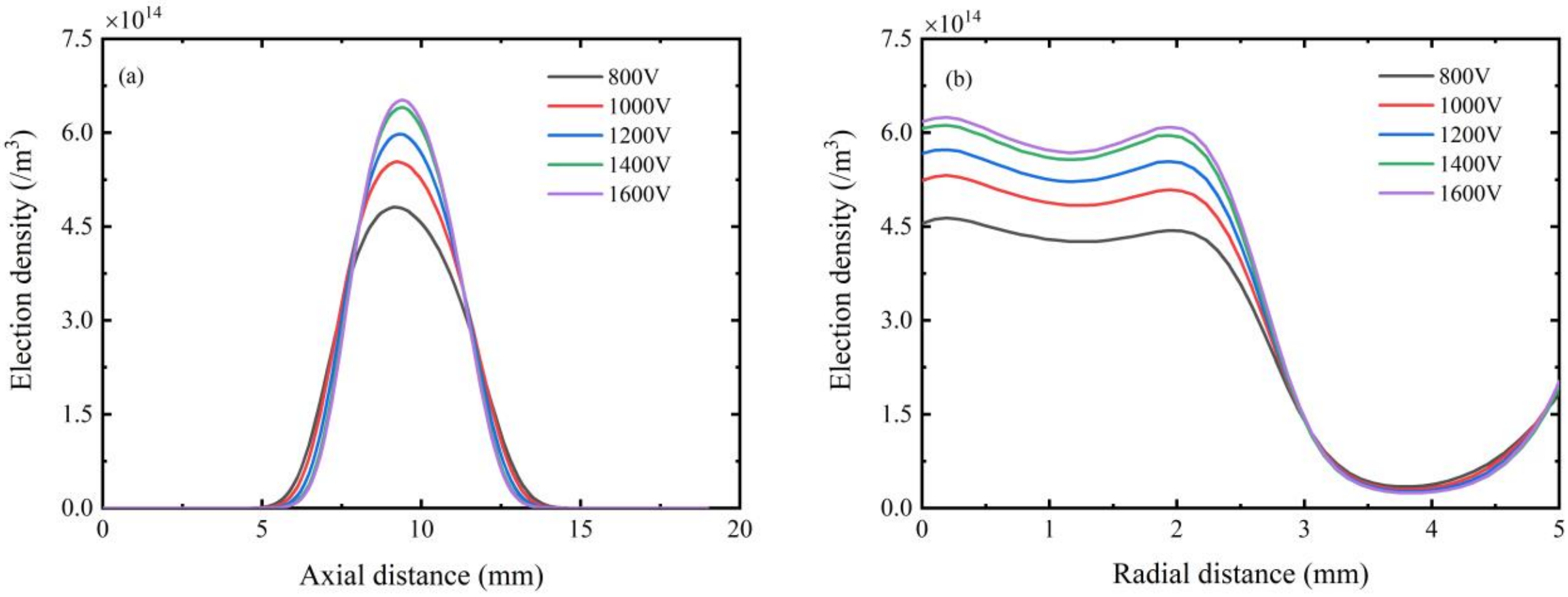

**Fig. 19** Distribution pattern of electron number density in axial and radial directions.

By varying the voltage applied to the anode tube, the distribution of hydrogen ion density profiles and the axial and radial hydrogen ion density curves at the center of the profile were obtained, as shown in **Figures 20 and 21**. As the anode voltage increased from 800 V to 1400 V, the peak axial hydrogen ion number density rose, while the radial hydrogen ion density gradually increased starting from the axial origin. The position corresponding to the peak moves closer to the anode plate as the anode voltage increases. When the voltage continues to increase to 1600 V, a decrease occurs in both cases. An increase in voltage typically implies a strengthening of the sheath electric field and a rise in the plasma potential. This may lead to two key effects: first, the enhanced electric field accelerates the movement of hydrogen ions toward the anode, resulting in faster loss, i.e., an increased ion loss rate; second, the overall rise in plasma potential may alter the equilibrium state of the discharge, reducing ionization efficiency or changing the electron energy distribution function, which is detrimental to the effective ionization of hydrogen.

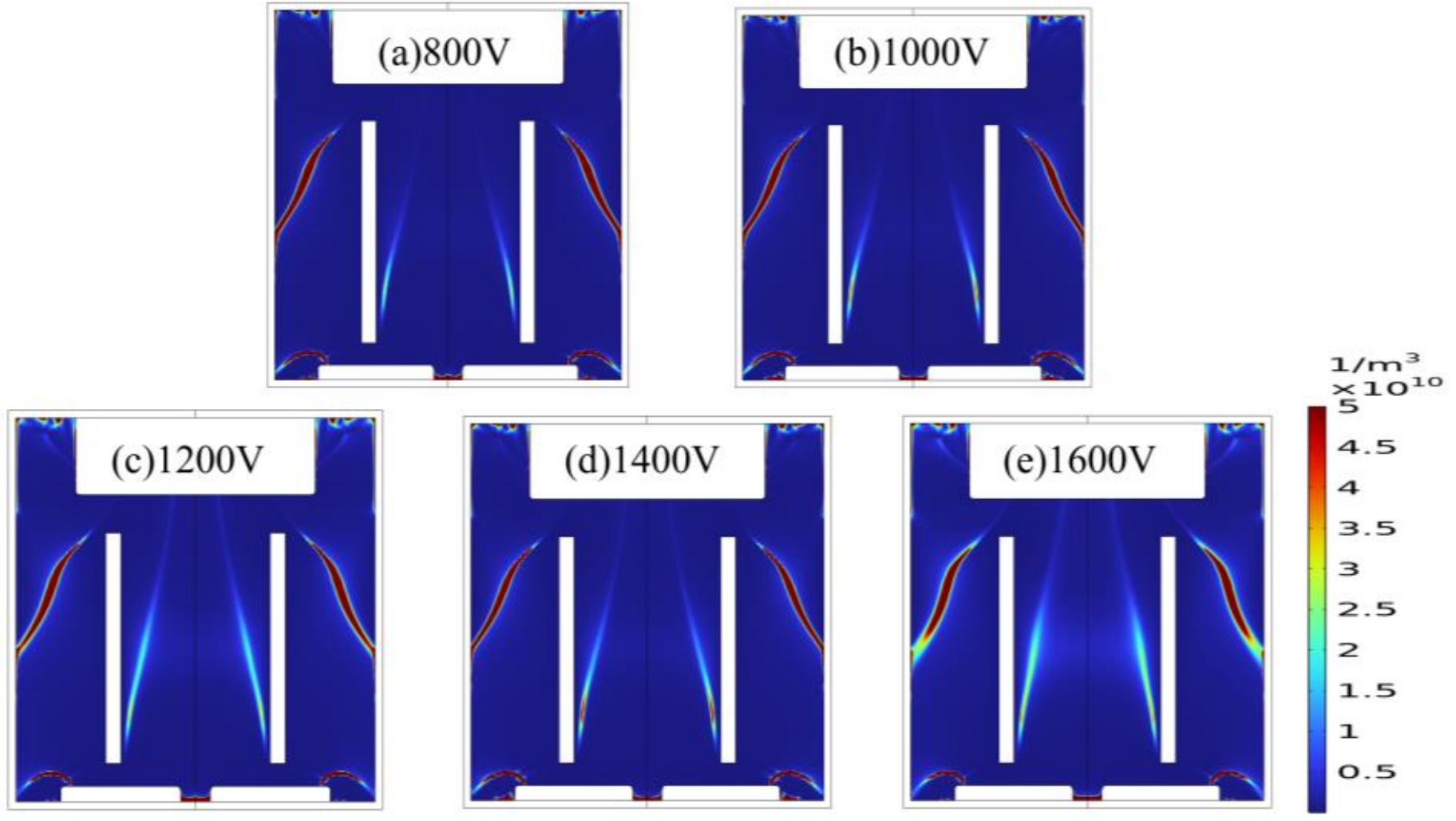

**Fig. 20** $H^+$ number density distribution map.

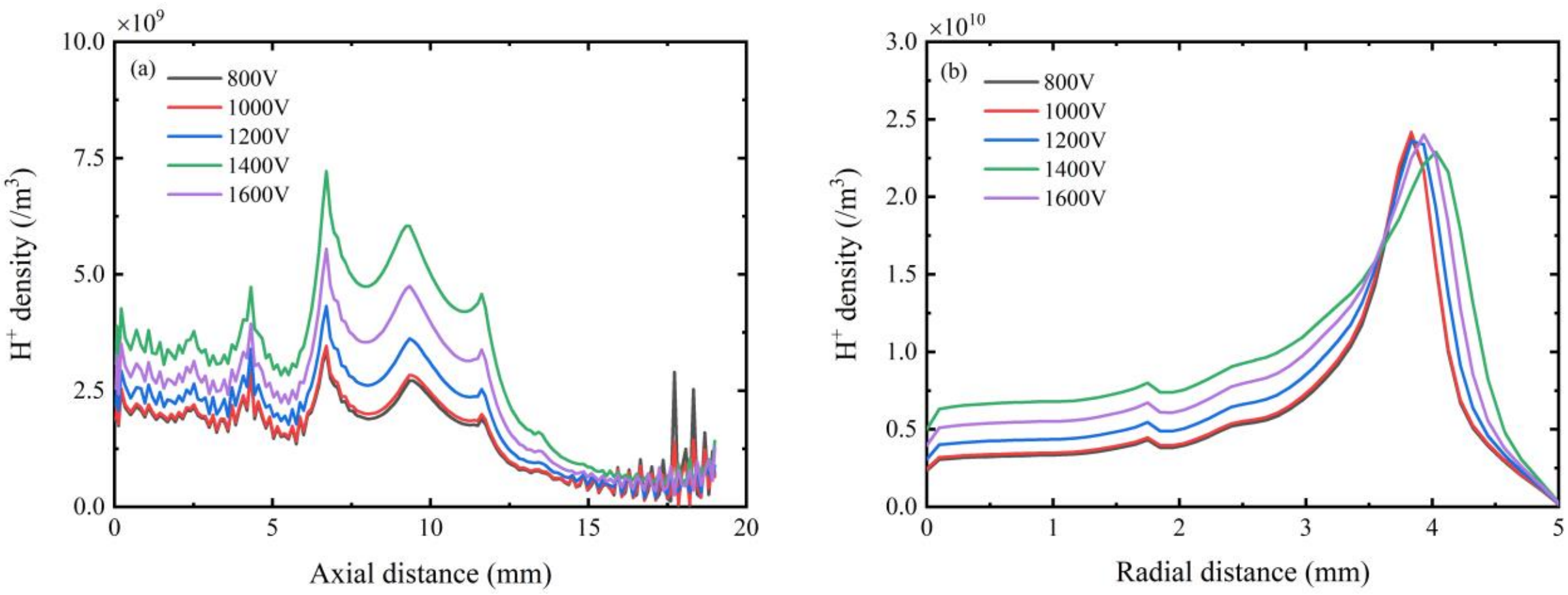

**Fig. 21** Distribution pattern of $H^+$ number density in axial and radial directions.

By varying the voltage applied to the anode tube, the distribution of hydrogen molecular ion number density was obtained, as shown in **Figure 22**. The distribution patterns of hydrogen molecular ion number density in the axial and radial directions are shown in **Figure 23**. The number density of hydrogen ions is inversely proportional to the anode voltage. Along the axial direction, the hydrogen molecular ion density decreases as the local anode voltage increases. This quantitative relationship reveals the important role of the voltage parameter in controlling the spatial distribution of plasma chemical composition.

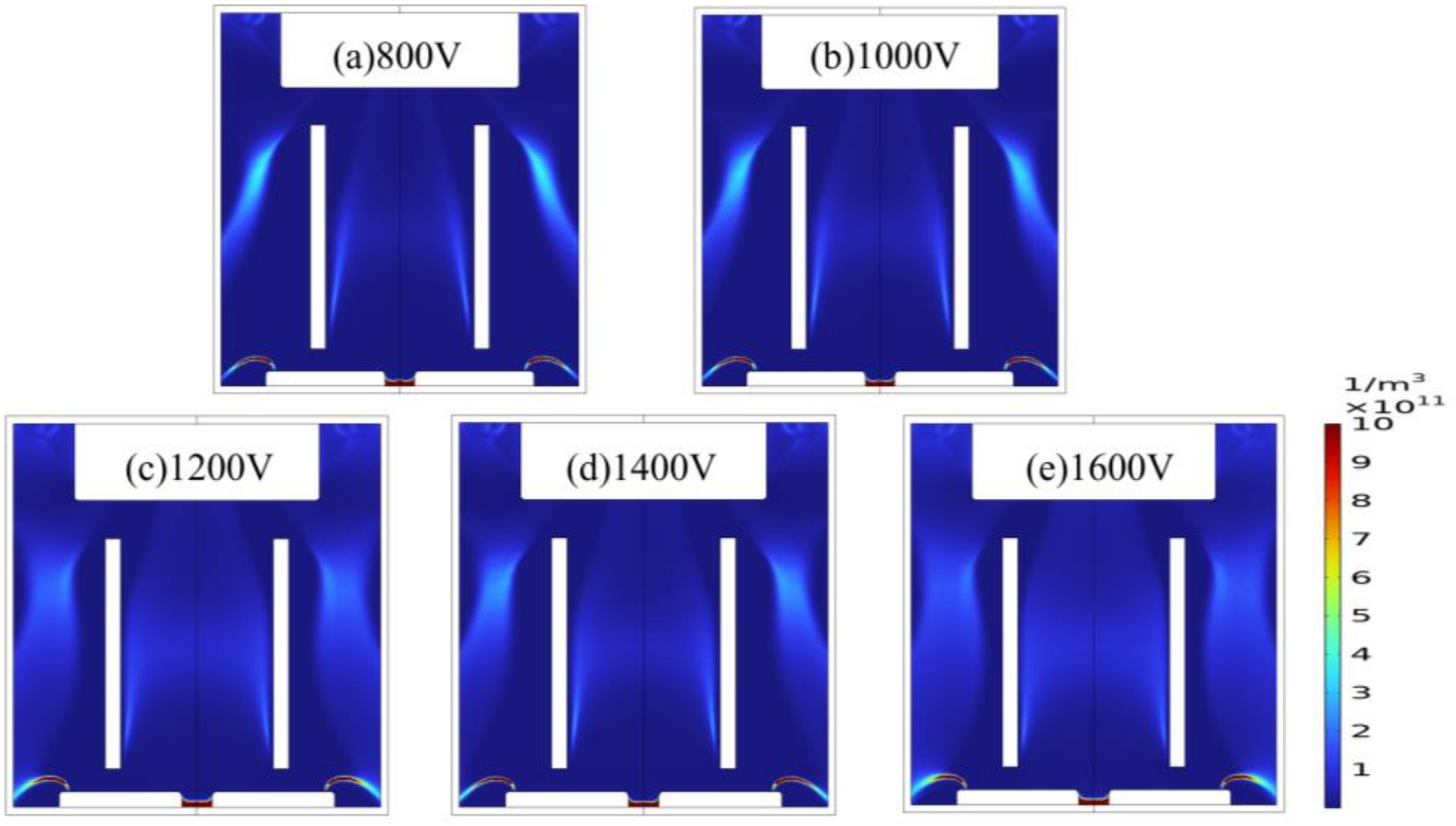

**Fig. 22** $H_2^+$ number density distribution map.

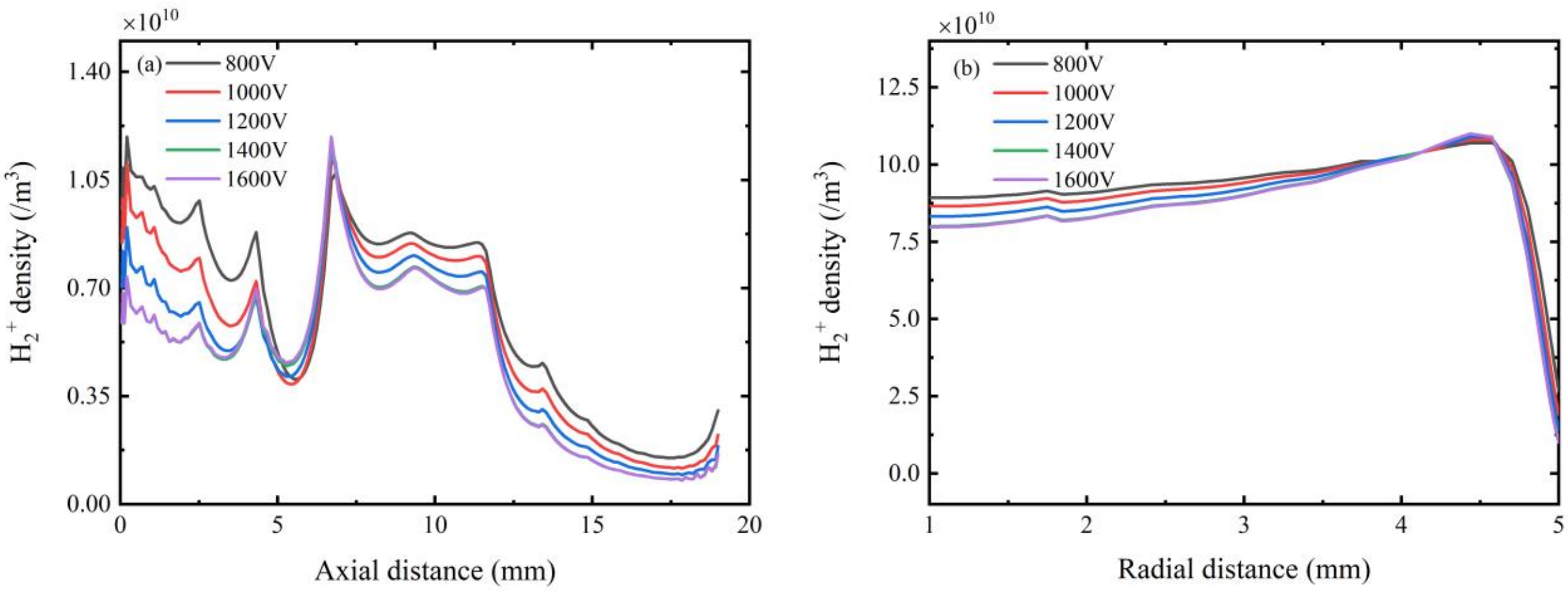

**Fig. 23** Distribution pattern of $H_2^+$ number density in axial and radial directions.

Building on this research, this paper further calculated the proportion of hydrogen monoatomic ions at different anode voltages, with the results shown in **Figure 24**. It should be noted that due to computational resource limitations and the fact that the simulation did not cover all reaction pathways, the results regarding the variation of the monoatomic ion ratio with anode voltage after discharge stabilization exhibit some deviation from actual conditions. Nevertheless, a clear pattern can still be observed in the figure: under hydrogen conditions with a fixed gas pressure of 0.5 Pa, the proportion of monoatomic ions generally increases initially and then decreases as the anode voltage increases. When the voltage on the anode cylinder reaches 1400 V, the proportion of monoatomic ions reaches a maximum of 30%.

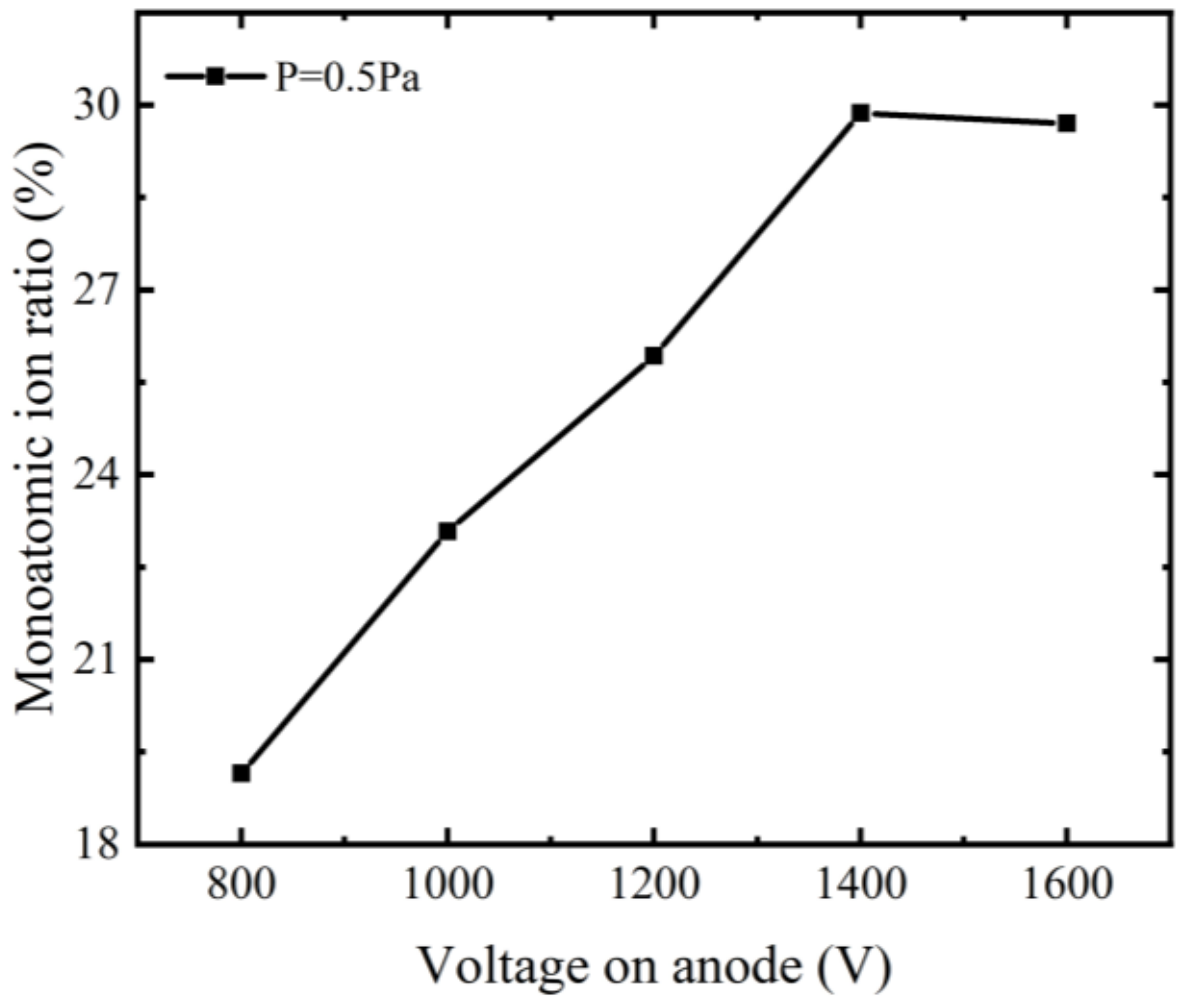

**Fig. 24** The ratio of the voltage on the anode cylinder to the monatomic ion.

## 4. Conclusion

This study addresses critical performance bottlenecks in sealed neutron tube Penning ion sources, including uneven magnetic field distribution, thermally induced demagnetization of permanent magnets, and low single-atom ion ratios, by completing soft iron-enhanced magnetic field structure design and plasma discharge parameter optimization. Through COMSOL multiphysics coupling simulations, we systematically elucidated the magnetic field concentration and homogenization mechanisms of soft iron structures, clarifying how working gas pressure and anode voltage regulate electron density, ion composition, and discharge stability. Results demonstrate that soft iron structures significantly enhance axial magnetic field intensity and uniformity in discharge regions, suppress permanent magnet demagnetization, and improve plasma confinement efficiency. Under optimal parameters of 0.06 Pa gas pressure and 1500 V anode voltage, the single-atom ion ratio in the ion source increased dramatically from approximately 9% in conventional structures to 30%, with notable improvements in ion beam quality and discharge efficiency. The proposed magnetic field configuration optimization and discharge parameter matching methodology provides reliable theoretical foundations and technical support for designing, developing, and engineering high-performance, long-life, and compact sealed neutron tube ion sources. This research holds significant value for advancing the practical application of compact neutron sources in industrial inspection, safety verification, and scientific research fields.

### CRediT authorship contribution statement

**Jia Song:** Conceptualization, Methodology, Software, Data curation, Validation, Investigation, Formal analysis, Writing-original draft. **Shengduo Liu:** Writing-review & editing, Software, Visualization. **Weiyang Zhang:** Resources. **Sijia Zhou:** Software, Visualization. **Hailong Xu:** Software, Visualization. **Zebin Li:** Formal analysis. **Tian Zhang:** Formal analysis. **Zhihua Gao:** Formal analysis. **Shiwei Jing:** Resources, Funding acquisition, Writing-review & editing, Conceptualization, Project administration. **Guofeng Qu:** Funding acquisition, Resources.

## Acknowledgments

This research was supported by the Education Department of Jilin Province of China (Grant No. JJKH20250303BS)